\documentclass[12 pt, article]{article}

\usepackage[left=1in, right=1in, top=0.75in, bottom=0.75in]{geometry}
\usepackage[utf8]{inputenc}
\usepackage{setspace}
\usepackage{graphicx}
\usepackage{authblk}
\usepackage{sectsty}
\usepackage{gensymb}
\usepackage{parskip}
\usepackage{caption}
\usepackage{titlesec}
\usepackage{wrapfig}
\usepackage[utf8]{inputenc}

\usepackage{xr}
\makeatletter

\newcommand*{\addFileDependency}[1]{% argument=file name and extension
\typeout{(#1)}% latexmk will find this if $recorder=0
\@addtofilelist{#1}
\IfFileExists{#1}{}{\typeout{No file #1.}}
}\makeatother

\graphicspath{{images/}}

\sectionfont{\fontsize{13}{16}\selectfont}

\usepackage[backend=biber, style=chem-acs,articletitle=true,sorting=none]{biblatex}
\titlespacing\section{0pt}{12pt plus 2pt minus 2pt}{12pt plus 2pt minus 2pt}
\title{\Large\textbf {Modulating the electrochemical intercalation of graphene interfaces with $\alpha$-RuCl$_3$ as a solid-state electron acceptor}}

\author[1]{Jonathon Nessralla}
\author[2]{Daniel T. Larson}
\author[3]{Takashi Taniguchi}
\author[4]{Kenji Watanabe}
\author[2,5]{Efthimios Kaxiras}
\author[1,6,*]{D. Kwabena Bediako}

\affil[1]{\textit{Department of Chemistry, University of California, Berkeley, CA 94720, USA}}
\affil[2]{\textit{Department of Physics, Harvard University, Cambridge, MA 02138, USA}}
\affil[3]{\textit{Research Center for Functional Materials, National Institute for Materials Science, Tsukuba 305-0044, Japan}}
\affil[4]{\textit{International Center for Materials Nanoarchitectonics, National Institute for Materials Science, Tsukuba 305-0044, Japan}}
\affil[5]{\textit{John A. Paulson School of Engineering and Applied Sciences, Harvard University, Cambridge, MA 02138, USA}}
\affil[6]{\textit{Chemical Sciences Division, Lawrence Berkeley National Laboratory, Berkeley, CA 94720, USA}}
\affil[*]{Correspondence to: bediako@berkeley.edu}

\date{}

\begin{document}
\maketitle
\doublespacing

\section*{\large{Abstract}}
Intercalation reactions modify the charge density in van der Waals (vdW) materials through coupled electronic--ionic charge accumulation, and are susceptible to modulation by interlayer hybridization in vdW heterostructures. Here, we demonstrate that charge transfer between graphene and $\alpha$-RuCl$_3$, which dopes the graphene positively, greatly favors the intercalation of lithium ions into graphene-based vdW heterostructures. We systematically tune this effect on Li$^+$ ion intercalation, modulating the intercalation potential, by using varying thicknesses of hexagonal boron nitride (hBN) as spacer layers between graphene and $\alpha$-RuCl$_3$. Confocal Raman spectroscopy and electronic transport measurements are used to monitor electrochemical intercalation and density functional theory computations help quantify charge transfer to both $\alpha$-RuCl$_3$ and graphene upon Li intercalation. This work demonstrates a versatile approach for systematically modulating the electrochemical intercalation behavior of two-dimensional layers akin to electron donating/withdrawing substituent effects used to tune molecular redox potentials.

\newpage

\section*{\large{Introduction}}
Layered van der Waals (vdW) materials are amenable to modulations in electronic and ionic charge accumulation through the intercalation of ions between their layers,\supercite{levy2012intercalated,whittingham1976intercalation} an integral process for electrochemical energy storage.\supercite{armand2008building,goodenough2013battery,simon2014batteries,maier2013thermodynamics} Thinning vdW materials to their two-dimensional (2D), monolayer (ML) limit and combining them with other vdW materials creates heterostructures with atomically sharp interfaces and varying electrochemical, electronic, and optical properties.\supercite{geim2013van,chaves2020bandgap,sun2015phosphorene,oakes2016interface} Previous studies investigated interface effects on the electrochemical intercalation of lithium ions between graphene, hexagonal boron nitride (hBN), and Mo$Ch_2$ ($Ch$ = S, Se).\supercite{zhao2018grhbn,bediako2018heterointerface,yang2022heterointerface} These studies revealed that graphene/Mo$Ch_2$ interfaces intercalate an order of magnitude more lithium than Mo$Ch_2$/Mo$Ch_2$ interfaces, and this intercalation reaction takes place at more positive potentials than the intercalation potential of few-layer graphene or graphene/hBN interfaces.

Here, taking inspiration from molecular approaches to control reduction potentials with electron donating/withdrawing substituents,\supercite{Kadish1976solvent,GodskLarsen2001substituent,Solis2011substituent} we sought to explore another route to modify electrochemical intercalation in 2D layered systems. The Mott insulator $\alpha$-RuCl$_3$, a vdW material with a deep work function ($\phi_{RuCl_3} =$ 6.1 eV),\supercite{Pollini1996Electronic,koitzsch2016Description} has been shown to hole-dope graphene ($\phi_{Gr} =$ 4.6 eV)\supercite{wang2020modulation} to a carrier density of $\sim 2-4\times10^{13}$ holes/cm$^2$.\supercite{wang2020modulation,mashhadi2019spinsplit,zhou2019evidence, rizzo2020polaritons} In this work, we use $\alpha$-RuCl$_3$ as a strong electron withdrawing layer to modulate the electrochemical intercalation of graphene with Li$^+$ ions. We tailor the electrochemical intercalation behavior by varying the distance between $\alpha$-RuCl$_3$ and graphene with hBN spacers to tune this electron withdrawing effect.

\section*{\large{Results and Discussion}}
\section*{Monitoring intercalation using Raman spectroscopy}

\begin{figure*}[tbhp]
    \centerline{\includegraphics[width=160mm]{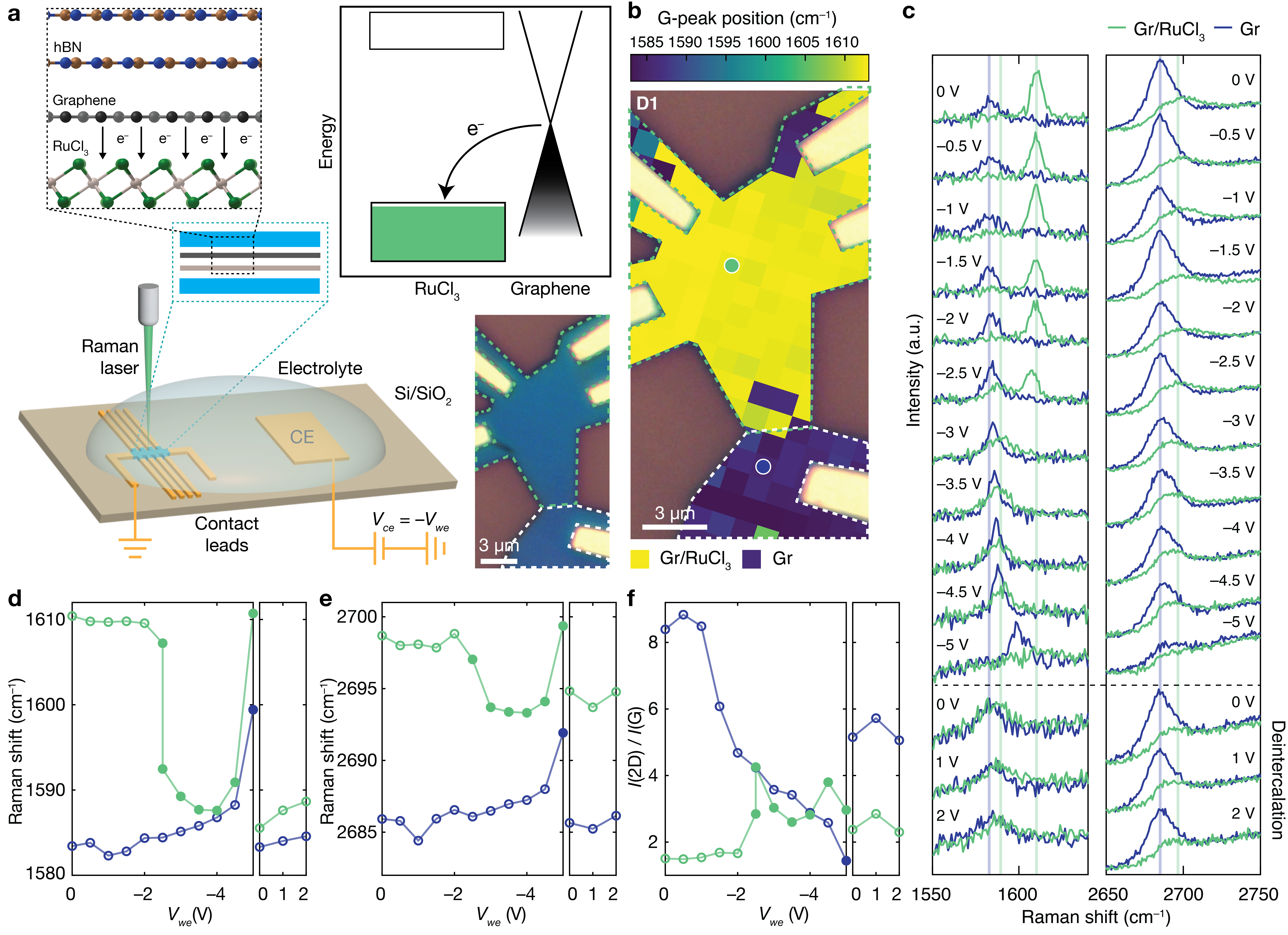}}
    \caption{\textbf{On-chip electrochemical intercalation of graphene interfaces monitored with Raman spectroscopy.} \textbf{(a)} Top-left, scheme of charge transfer across the graphene/$\alpha$-RuCl$_3$ interface. Top-right, scheme of relative band alignments and work function differences between graphene and $\alpha$-RuCl$_3$. Bottom-left, scheme of the on-chip electrochemical cell. Bottom-right, optical micrograph of D1; dashed lines distinguish the Gr (white) and GrRu (green) regions. \textbf{(b)} Confocal Raman map of the graphene G peak position of device D1 before intercalation. Circles indicate locations associated with the spectra in \textbf{c}. \textbf{(c)} Raman spectra of the graphene G (left panel) and 2D (right panel) peaks as $V_{we}$ (numbers above each spectrum) is swept for D1. The vertical bars are added to guide the eye. \textbf{(d,e)} G and 2D peak position as a function of $V_{we}$, respectively, for the spectra shown in \textbf{c}. Open and closed circles are used to indicate our interpretation of non-intercalated and intercalated, respectively. \textbf{(f)} Intensity ratio of the 2D to G peak as a function of $V_{we}$ for the spectra shown in \textbf{c}.}
    \label{fig.raman}
\end{figure*}

To compare the intercalation of lithium ions into different graphene interfaces, heterostructures of hBN, graphene, and $\alpha$-RuCl$_3$ were constructed by mechanical exfoliation of each vdW material\supercite{huang2015reliable} and subsequent stacking of the thinned crystals using a standard dry transfer method.\supercite{frisenda2018deterministic,wang2013edgecontact} The first device (D1) consists of a top and bottom hBN that fully encapsulates a continuous monolayer (ML) graphene sheet interfaced partially with ML $\alpha$-RuCl$_3$. This partial interfacing yields a single device possessing two regions that can be probed simultaneously: an hBN/graphene/hBN (Gr) region as well as an hBN/graphene/$\alpha$-RuCl$_3$/hBN (GrRu) region. The Gr region therefore serves as an internal reference and point of comparison for electrochemical changes induced by $\alpha$-RuCl$_3$. The Raman map of D1 shown in Figure \ref{fig.raman}b reveals a doped graphene region with $\alpha$-RuCl$_3$ and a pristine, undoped graphene region without it, consistent with previous work.\supercite{wang2020modulation,mashhadi2019spinsplit,zhou2019evidence}

An on-chip electrochemical cell was fabricated by patterning a Pt counter electrode (electrolyte gate) and electrically contacting the graphene heterostructure to use as the so-called `working electrode'. The electrochemical intercalation of these systems was probed by sweeping the counter electrode potential ($V_{ce}$) to more positive potentials incrementally (corresponding to increasingly negative working electrode potentials, $V_{we}$, where $V_{we} = -V_{ce}$) and monitoring changes of the graphene-based Raman spectral features.

% Figure 1 %

Figure \ref{fig.raman}c shows the Raman spectra of the graphene G and 2D peaks in D1 as the electrochemical bias is varied. Figures \ref{fig.raman}d--f display changes in the G and 2D peak positions and heights of these spectra as a function of $V_{we}$. For the GrRu region, there is a noticeable discontinuity in the G peak position, the 2D peak position, and the 2D/G intensity ratio between $-2.5$ and $-3$ V that is consistent with a significant change in doping and/or strain based on previous work.\supercite{lee2012optical,wang2020modulation,das2008monitoring,pisana2007breakdown,yan2007electric} We interpret this as a sign of a large influx of lithium ions. For the Gr region, the changes seen in the G and 2D peak positions and the 2D/G peak height ratio are consistent with merely electrostatic gating from the electrolyte or dilute intercalation until about $-4.5$ V.\supercite{zhao2018grhbn,bediako2018heterointerface} At $-5$ V, there is a sudden blue shift in the position of the G and 2D peaks, and a relatively large drop in the 2D/G height ratio, which is consistent with the Gr region undergoing an intercalation event with a large influx of lithium ions.\supercite{zhao2018grhbn,bediako2018heterointerface} Taken together, this analysis of the Raman spectra suggests that the onset potential for intercalation has shifted by at least $+2.5$ V (compared to a $+0.5$ V shift for graphene/Mo$Ch_2$ interfaces)\supercite{bediako2018heterointerface,yang2022heterointerface} with the introduction of $\alpha$-RuCl$_3$ into the graphene heterostructure.

After the measurement at $-5$ V, the $V_{we}$ is swept to 0 V and D1 is held at this voltage for $\sim$3 hours to drive the lithium ions out of the heterostructure. In the GrRu region, there is no recovery of the original G peak position, even upon applying an opposite bias of $V_{we} = +2$ V. This result implies that insertion of lithium ions into this region of the heterostructure may be partially irreversible. For the Gr region, the G peak largely returns to the original position by $V_{we} = 0$, indicating effectively full deintercalation in this region. While Raman spectroscopy provides helpful insight into the intercalation/deintercalation of lithium into graphene-containing heterostructures, these measurements do not explicitly reveal whether the graphene is electron or hole doped. To more precisely monitor the changes in carrier density and to interrogate changes in carrier type as the intercalation reaction progresses, we turn to electronic transport measurements.\supercite{Kuhne2017ultrafast,bediako2018heterointerface,zhao2018grhbn}

\section*{\textit{Operando} electronic transport}

\begin{figure*}[tbh]
    \centerline{\includegraphics[width=86mm]{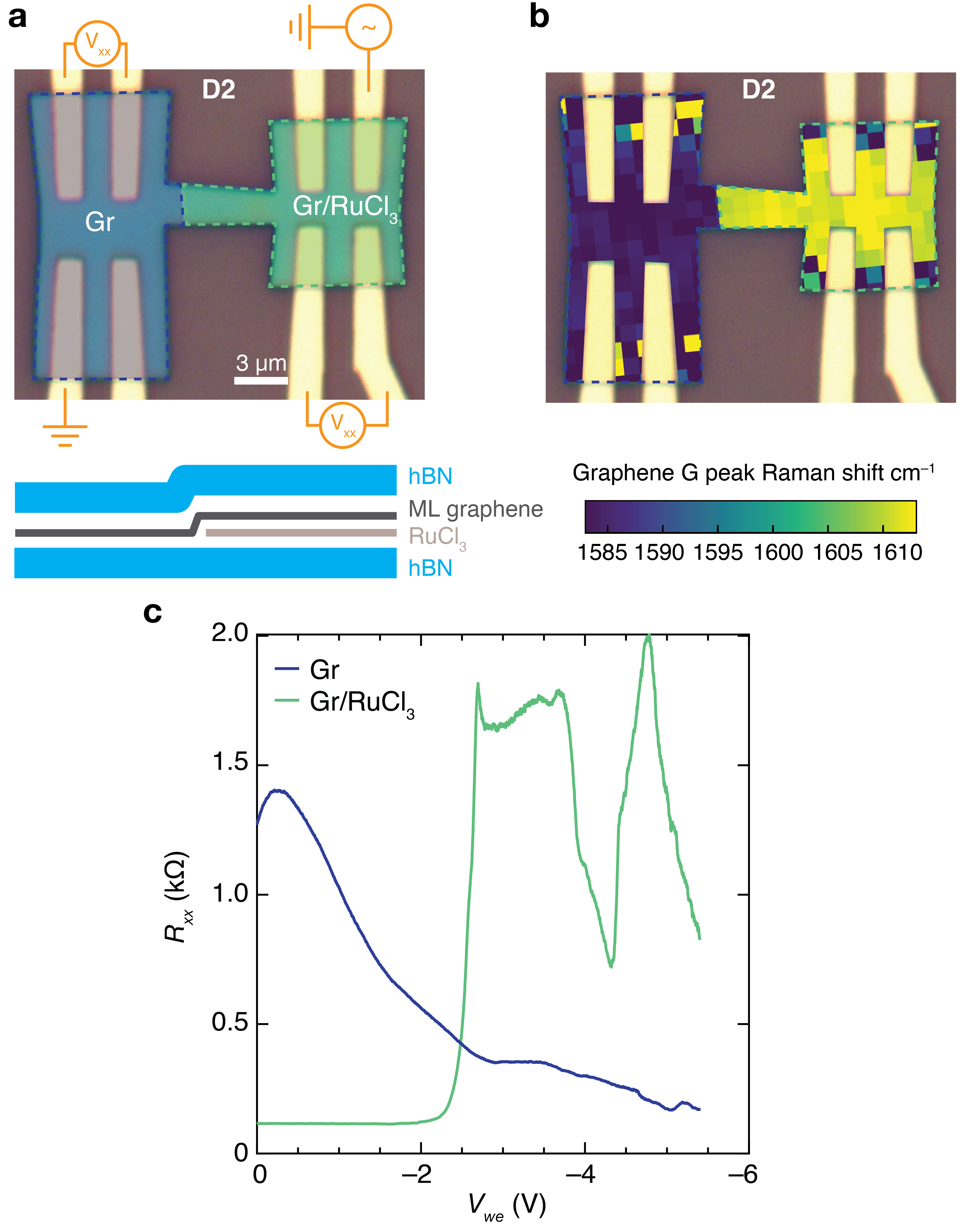}}
    \caption{\textbf{On-chip electrochemical intercalation monitored with electronic transport measurements.} \textbf{(a)} Top: Optical micrograph of the device used for \emph{in situ} transport measurements (D2). The Gr and GrRu regions are shown with false colors overlaid. The schematic of the circuit setup shows how the contact leads are used to obtain longitudinal ($V_{xx}$) voltage measurements. Bottom: Cross-sectional schematic of D2. \textbf{(b)} Raman map of the graphene G peak shift of D2 before intercalation. \textbf{(c)} Longitudinal resistance ($R_{xx}$) for both regions as $V_{we}$ is swept.}
    \label{fig.transport}
\end{figure*}

Measurements of longitudinal ($R_{xx}$) resistance have been used previously to continuously monitor intercalation reactions in graphene heterostructures.\supercite{Kuhne2017ultrafast,zhao2018grhbn,bediako2018heterointerface} The $R_{xx}$ measurement gives insight into the intercalation progress in graphene since $R_{xx}$ is inversely proportional to both carrier density and mobility. Lithium intercalation increases the number of electron carriers in graphene and decreases its mobility as the lithium ions act as scattering sites when coupled to the graphene lattice.\supercite{Kuhne2017ultrafast,zhao2018grhbn} The device used (D2) for the measurements in this section follows the same device scheme and fabrication procedure as D1. D2 contains a Gr and GrRu region, and an optical micrograph of D2 is shown in Figure \ref{fig.transport}a.
Figure \ref{fig.transport}b shows the Raman map of device D2 before intercalation, revealing two distinctly doped regions with G peak positions consistent with those observed in device D1 for Gr and GrRu regions.

% Figure 2 %

Figure \ref{fig.transport}c shows $R_{xx}$ as a function of $V_{we}$ for both the Gr and GrRu regions in D2. For the Gr region, the trends in the $R_{xx}$ data are consistent with what is reported previously for graphene/hBN heterostructures under an increasing electrochemical bias.\supercite{zhao2018grhbn,bediako2018heterointerface} No spike in $R_{xx}$ from a large influx of lithium ions is observed up to $-5.5$ V. We therefore assign the onset intercalation potential for the Gr region of D2 as beyond $-5$ V (consistent with previous results with graphene only devices\supercite{zhao2018grhbn,bediako2018heterointerface}). 

For the GrRu region, there is a jump in $R_{xx}$ that peaks at $-2.7$ V. This spike in $R_{xx}$ is consistent with a rapid decrease of hole carriers as the Fermi level moves towards the charge neutrality point of graphene during Li$^+$ intercalation in addition to any decrease in mobility due to ion insertion into the heterostructure. This level of sudden doping in graphene is not expected from pure electrostatic gating, but it is expected from the intercalation of lithium ions. From these data, we can tentatively assign $-2.5$ V as the onset intercalation potential for the GrRu region of D2.
Overall, these transport data are consistent with the results of our \emph{in situ} Raman data: the onset of intercalation in GrRu appears shifted by more than $+2.5$ V relative to the Gr region. Next, we precisely quantify the number and type of carriers in these heterostructures throughout intercalation. We also modulate the doping induced by $\alpha$-RuCl$_3$ with hBN spacer layers to further tune the intercalation onset potential. 

\section*{Modulation of doping and intercalation with hBN spacers}

\begin{figure*}[tbhp]
    \centerline{\includegraphics[width=175mm]{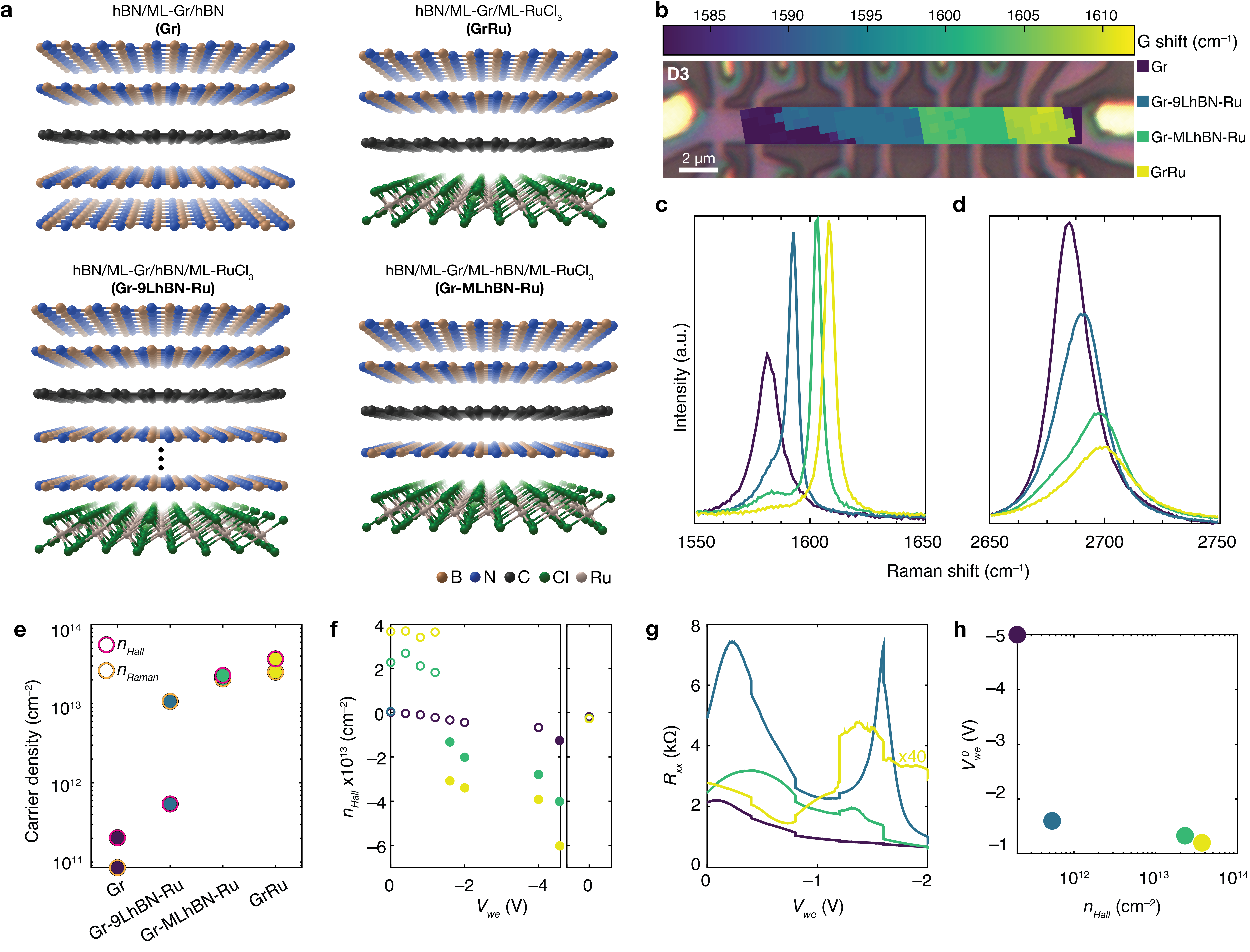}}
    \caption{\textbf{Modulation of intercalation with hBN spacers.} \textbf{(a)} Illustrations of the various regions of D3. \textbf{(b)} Raman map of the graphene G peak shift overlaid over an optical micrograph of D3. \textbf{(c,d)} Average Raman spectra over each device region in (b) in the G, \textbf{c}, and 2D, \textbf{d}, spectral range. \textbf{(e)} Determination of carrier density from the Raman data ($n_{Raman}$) and Hall field sweeps ($n_{Hall}$) in each region before intercalation. \textbf{(f)} Carrier density of each region as a function of $V_{we}$ as calculated using the Hall effect measurements ($n_{Hall}$). Open circles indicate carrier densities prior to the onset of intercalation and filled circles indicate a region is beyond the onset potential (indicated by a jump in $n_{Hall}$). \textbf{(g)} $R_{xx}$ of each region as a function of $V_{we}$. $R_{xx}$ for the GrRu region has been multiplied by a factor of 40 for clarity. \textbf{(h)} Onset intercalation potential ($V_{we}^0$) for each region shown as a function of the initial doping in that region.}
    \label{fig.hbn}
\end{figure*}

Previous studies have shown that the $p$-doping induced in graphene by proximal $\alpha$-RuCl$_3$ can be attenuated by increasing the distance between the graphene and the $\alpha$-RuCl$_3$ layers,\supercite{wang2020modulation} using hBN sheets of varying thicknesses. To directly compare the electrochemical behavior as the doping from the interfacial charge transfer is modified, we fabricated a single heterostructure with continuous sheets of ML graphene and ML $\alpha$-RuCl$_3$ divided into regions with different thicknesses of hBN spacers between. Figure \ref{fig.hbn}a depicts the structures associated with different regions in D3: hBN/ML graphene/hBN (Gr); hBN/ML graphene/ML $\alpha$-RuCl$_3$ (GrRu); hBN/ML graphene/$\sim$9 layer hBN/ML $\alpha$-RuCl$_3$ (Gr-9LhBN-Ru); hBN/ML graphene/ML hBN/ML $\alpha$-RuCl$_3$ (Gr-MLhBN-Ru).

% Figure 3 %

The initial doping of each region in D3 is established with Raman mapping in Figure \ref{fig.hbn}b. Figure \ref{fig.hbn}c and d show the G and 2D peak spectral regions, respectively, for each region. An approximation of the initial doping by region, (Figure \ref{fig.hbn}e), can be made using the position of the G peak.\supercite{das2008monitoring,pisana2007breakdown,yan2007electric} This calculation can be compared to the carrier density obtained from the Hall effect. The two methods of estimation are consistent for each region except the Gr-9LhBN-Ru region. This could be due to the fact that the G peak position is not only correlated to doping in graphene, but also the strain.\supercite{lee2012optical}

In D2, electrical leads continuously probe $R_{xx}$ of both the Gr and GrRu regions as $V_{we}$ is swept. For the intercalation of device D3, $V_{we}$ is swept to a specified value at $\sim$330 K and then rapidly cooled to 200 K to suspend the intercalation progress. Then the carrier density is measured by sweeping the external magnetic field, $B$, and recording $R_{xy}(B)$. Figure \ref{fig.hbn}f shows the Hall carrier density measured in this way at each increment of $V_{we}$. Notably, the GrRu and Gr-MLhBN-Ru regions switch precipitously from $>10^{13}$ holes/cm$^2$ to $>10^{13}$ electrons/cm$^2$ between $-1.2$ and $-1.6$ V, and this very large increase in electron-doping indicates that Li$^+$ intercalation has occurred. In contrast, the Gr region only gradually accumulates electrons and does not display any rapid change in carriers until about $-5$ V. We note that we were unable to measure the $R_{xy}(B)$ for the Gr-9LhBN-Ru region because of contact failure. As in the case of devices D1 and D2, device D3 therefore shows a $>$2.5 V change in the onset intercalation potential for both the GrRu and Gr-MLhBN-Ru regions relative to the Gr region.

Figure \ref{fig.hbn}g shows $R_{xx}$ for each region as a function of $V_{we}$. While this measurement tracks $R_{xx}$ continuously as $V_{we}$ is swept, there are discontinuities every 0.4 V to conduct Hall measurements. Despite these discontinuities, we can still observe significant increases in $R_{xx}$ that are in line with intercalation as indicated by the $R_{xy}$ data. Together, Figures \ref{fig.hbn}f and g allow us to estimate $-1.2$ V, $-1.3$ V, and $-1.6$ V as the onset intercalation potentials for the GrRu, Gr-MLhBN-Ru, and Gr-9LhBN-Ru regions, respectively. This result is consistent with the spectroscopic data of device D1 and transport data of D2; the hole doping and increased work function decrease the onset potential for lithium ion insertion into the vdW interface. By having multiple regions in one device (D3), we are able to show that the intercalation onset can be systematically modulated by tuning the doping induced by the $\alpha$-RuCl$_3$ with hBN spacers (see Figure \ref{fig.hbn}h).  

% Figure 4 %
\begin{figure*}[tbhp]
    \centerline{\includegraphics[width=160mm]{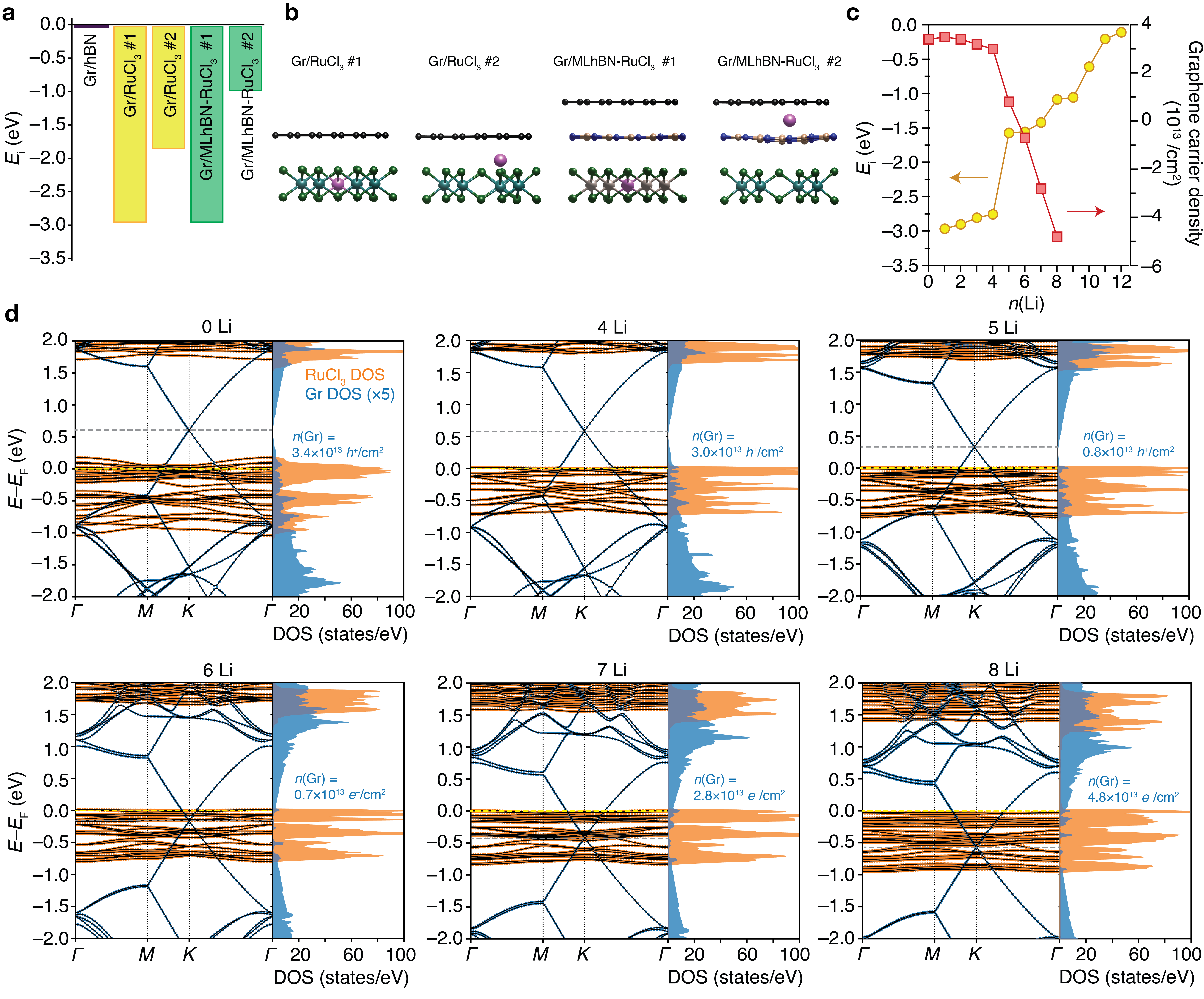}}
    \caption{\textbf{Computed intercalation energies and band structures.} \textbf{(a)} Computed intercalation energies for 1 Li atom in Gr/hBN, Gr/RuCl$_3$, and Gr/MLhBN/RuCl$_3$ heterostructures. \textbf{(b)} Structures of Li intercalated heterostructures corresponding to the energies in (a). In the case of Gr/RuCl$_3$ we depict the most favorable location (1) and second most favorable location (2). For Gr/MLhBN/RuCl$_3$, we show the most favorable location for each interface. \textbf{(c)} Intercalation energies (circles) and computed graphene carrier densities (squares) for sequential addition of Li atoms to Gr/RuCl$_3$ heterostructures. \textbf{(d)} Computed band structures and DOS for selected amounts of Li intercalated into Gr/RuCl$_3$ heterostructures. Graphene-based bands and DOS are depicted in blue and $\alpha$-RuCl$_3$-based bands and DOS are depicted in orange. In all cases, the graphene DOS plots have been multiplied by 5 for visibility. The Fermi level, $E-E_F=0$ is indicated by a horizontal dashed yellow line and the energy of the charge neutrality point of graphene is indicated by a horizontal dashed grey line.} 
    \label{fig.scheme}
\end{figure*}

\section*{Theoretical computations of Li intercalation}

The impact of $\alpha$-RuCl$_3$ on Li intercalation energetics can be further understood from density functional theory (DFT) computations. First, we estimate the theoretical Li binding energies for intercalation at various interfaces in Gr, GrRu, and Gr-MLhBN-Ru heterostructures. As shown in Figure \ref{fig.scheme}a, we find that in a Gr/hBN heterostructure (no $\alpha$-RuCl$_3$), for Li situated above the hollow of a C$_6$ hexagon of graphene and under a N atom of hBN, the Li binding energy is computed to be $-0.033$ eV, indicating very slight energetic favorablity for Li intercalation. However, in a Gr/RuCl$_3$ heterostructure, the most energetically favorable location for Li is within the hollows of the $\alpha$-RuCl$_3$ lattice, in the $\alpha$-RuCl$_3$ plane. This location is calculated to possess a binding energy of $-2.97$ eV. In contrast, a Li atom positioned in the interlayer region between graphene and $\alpha$-RuCl$_3$ is computed to have a binding energy of $-1.87$ eV.

With an hBN monolayer between graphene and $\alpha$-RuCl$_3$ the Li binding energy at the Gr/hBN interface is strongly augmented to about $-1.0$ eV (from $-0.033$ eV), but the most energetically favorable location remains within the $\alpha$-RuCl$_3$ lattice with a binding energy of $-2.97$ eV (unchanged from the case where hBN is absent). These calculations indicate that the binding of Li in vdW heterostructures is very strongly favored by the large electron withdrawing ability of $\alpha$-RuCl$_3$, but also reveal that charge transfer to $\alpha$-RuCl$_3$ upon Li intercalation is very favorable. Figure \ref{fig.scheme}b shows the atomic models of these intercalated Li$^+$ ions in the aforementioned interfaces.

We next consider the modification of Li binding energies upon sequential intercalation of multiple Li atoms to a Gr/$\alpha$-RuCl$_3$ heterostructure. Figure \ref{fig.scheme}c shows the progression of Li binding energy as a function of the number of Li atoms added. Here, we compute in each case the most favorable location for each added Li. The first 4 Li atoms insert into the hollows of the $\alpha$-RuCl$_3$ lattice, while the next 8 are located between the Gr and $\alpha$-RuCl$_3$ layers. By the 12th Li atom, the Li intercalation energy ($-0.106$ eV) is computed to be approaching that of an unmodified Gr-hBN interface (Figure \ref{fig.scheme}a). 

Computed band structures and density of states (DOS) in Figure \ref{fig.scheme}d show the evolution of the relative positions of $\alpha$-RuCl$_3$ and Gr bands as Li is intercalated, explaining the previously discussed binding energies in the context of charge transfer to $\alpha$-RuCl$_3$ and graphene. Prior to intercalation, the calculated band structures are consistent with strong hole-doping of graphene by $\alpha$-RuCl$_3$. Intercalation of 4 Li atoms into the heterostructure (positioned within the hollows of the $\alpha$-RuCl$_3$ lattice) involves charge transfer to only $\alpha$-RuCl$_3$ and minimal doping of Gr. Subsequent addition of a 5th Li atom (in the interlayer region) dopes the Gr lattice, initiating a reversal of the hole doping of Gr by $\alpha$-RuCl$_3$. We find that intercalation of 8 Li atoms produces a doping level of almost 5 $\times$ $10^{13}$ electrons/cm$^2$, which is close to the value that is eventually accessed experimentally (Figure \ref{fig.hbn}f).

The Raman spectroscopy (Figure \ref{fig.raman}) and electronic transport (Figures \ref{fig.transport} and \ref{fig.hbn}) experimental results show that intercalation of GrRu heterostructures occurs precipitously around $V_{we}=-2$ V to produce an electron doped graphene. Thus, there are no experimental indications of an isolated intermediate state associated with exclusive intercalation and doping of $\alpha$-RuCl$_3$. We speculate that this may be rationalized by considering that charge balance requires that insertion of Li$^{+}$ ions into the $\alpha$-RuCl$_3$ lattice must be compensated by electron transport through the solid. As $\alpha$-RuCl$_3$ is an insulator, this coupled electronic--ionic insertion should be prohibitively slow if reliant on electronic transport through RuCl$_3$. Instead, we propose that intercalation of the heterostructure takes place only upon biasing the electrode to a potential at which intercalation of graphene becomes spontaneous. Still, upon Li$^+$ ion insertion, we expect substantial lithiation of $\alpha$-RuCl$_3$ and experimental evidence for charge transfer to $\alpha$-RuCl$_3$ upon Li intercalation is found upon deintercalation of the heterostructure (Figure \ref{fig.hbn}f). As discussed previously, after deintercalation, graphene is returned to a nearly charge-neutral state, and not to the initial strongly hole-doped state. This result reveals that Li intercalation and charge transfer into $\alpha$-RuCl$_3$ has taken place irreversibly and future work will be required to experimentally measure the amount of Li inserted into the $\alpha$-RuCl$_3$ lattice itself and to interrogate the proposed non-equilibrium intercalation mechanism.

\section*{\large{Conclusions}}
These results demonstrate that the strong hole doping induced from interfacing graphene with $\alpha$-RuCl$_3$ significantly reduces the magnitude of external electrochemical bias ($V_{we}$) needed to drive the intercalation of lithium ions, with the intercalation onset shifted by $>$2.5 V relative to the intercalation of undoped graphene that is only encapsulated by hBN. This decrease in potential needed for intercalation is shown to be tunable by inserting hBN spacers between the graphene and $\alpha$-RuCl$_3$ layers. Computations show that intercalation results in strong electron doping of both $\alpha$-RuCl$_3$ and graphene, but experimentally, intercalation appears to only be initiated when charge transfer to graphene becomes spontaneous. These results demonstrate a distinctive level of control for electronic and ionic charge accumulation in vdW heterostructures, which may spur future studies on tailoring intercalation (electro)chemistry in 2D materials.

\section*{\large{Associated Content}}
\textbf{Supporting Information.} Crystal growth and characterization, sample/device fabrication, Raman spectroscopy, details of electrical transport measurements and density functional theory computations.

\section*{\large{Author Information}}
\textbf{Corresponding Author} 

Correspondence and requests for materials should be emailed to DKB (email: bediako@berkeley.edu).

\textbf{Author Contributions} 

JNN and DKB conceived the study, JNN performed the experiments, DTL and EK performed the theoretical calculations, TT and KW provided the hBN crystals. JNN and DKB analyzed the data and wrote the manuscript with input from all co-authors.

\section*{\large{Acknowledgments}}
This material is based upon work supported by the US Department of Energy, Office of Science, Office of Basic Energy Sciences, under award no. DE-SC0021049 (DKB) and the Gordon and Betty Moore Foundation EPiQS Initiative Award no. 10637. This research made use of the computational resources of Frontera at the Texas Advanced Computing Center (DTL, EK). Frontera is made possible by National Science Foundation award OAC-1818253. Additional calculations were run on the FASRC Cannon cluster supported by the FAS Division of Science Research Computing Group at Harvard University. EK and DTL acknowledge funding from the STC Center for Integrated Quantum Materials, NSF Grant No. DMR-1231319; NSF Award No. DMR-1922172; the Army Research Office under Cooperative Agreement Number W911NF-21-2-0147; and the Simons Foundation, Award No. 896626. Confocal Raman spectroscopy was supported by a Defense University Research Instrumentation Program grant through the Office of Naval Research under award no. N00014-20-1-2599 (DKB). Other instrumentation used in this work was supported by grants from the W.M. Keck Foundation (Award no. 993922), and the 3M Foundation through the 3M Non-Tenured Faculty Award (no. 67507585). K.W. and T.T. acknowledge support from JSPS KAKENHI (Grant Numbers 19H05790, 20H00354 and 21H05233).

\section*{\large{Competing Interests}}
The authors declare no competing interests.

\printbibliography

\end{document}